\documentclass[12pt]{article}

\usepackage{graphicx}
\usepackage{bm}

\usepackage{SIunits}
\usepackage[superscript,biblabel]{cite}

\usepackage{hyperref}

\usepackage{color}

\usepackage{mathrsfs}
\usepackage{resizegather}
\everymath{\displaystyle}

\begin{document}
\baselineskip24pt

\title{}
{\noindent\large{\textbf {%
Dark-field second harmonic generation imaging of grain boundaries in 2D semiconducting transition metal dichalcogenides
}}}
\\
\author{}
Bruno R. Carvalho$^{1,\ast}$, Yuanxi Wang$^{2,3,4}$, Kazunori Fujisawa$^{2,3}$, Tianyi Zhang$^{3,5}$, Ethan Kahn$^{3,5}$, Ismail Bilgin$^{6}$, Pulickel M. Ajayan$^{7}$, Ana M. de Paula$^{8}$, Marcos A. Pimenta$^{8}$, Swastik Kar$^{6}$, Vincent H. Crespi$^{2,3,4}$, Mauricio Terrones$^{2,3,5}$ \& Leandro M. Malard$^{8,\ast}$
\\
\\
$^{1}$ Departamento de F\'isica, Universidade Federal do Rio Grande do Norte, Natal, Rio Grande do Norte 59078-970, Brazil \\
$^{2}$ Department of Physics, The Pennsylvania State University, University Park, PA 16802, USA \\
$^{3}$ Center for 2-Dimensional and Layered Materials, The Pennsylvania State University, University Park, PA 16802, USA \\
$^{4}$ 2-Dimensional Crystal Consortium, The Pennsylvania State University, University Park, PA 16802, USA \\
$^{5}$ Department of Materials Science and Engineering, The Pennsylvania State University, University Park, PA 16802, USA \\
$^{6}$ Department of Physics, Northeastern University, Boston, Massachusetts 02115, USA \\
$^{7}$ Department of Material Science and NanoEngineering, Rice University, Houston, TX, 77005, United States \\
$^{8}$ Departamento de F\'isica, Universidade Federal de Minas Gerais, Belo Horizonte, Minas Gerais 30123-970, Brazil \\
$^{\ast}$E-mail: brunorc@fisica.ufrn.br (B.R.C.) and lmalard@fisica.ufmg.br (L.M.M.).
\\


\newpage
\noindent{\textbf{
A material's {\it quality} often determines its {\it identity} at both a fundamental and applied level. For example, the fractional quantum Hall effect simply does not exist in heterostructures of insufficient crystalline quality~\cite{du2009,bolotin2009,dean2011}, while silicon wafers~\cite{atabaki2018,boettcher2010,garnett2010} or non-linear crystals~\cite{nikogosyan2005,morris1990} of low crystalline purity are electronically or optically useless. Even when surface states are topologically protected their visibility depends critically on obtaining extremely high crystalline quality in the bulk regions~\cite{wang2013}. Extended defects with one dimensionality smaller than that of the host, such as 2D grain boundaries in 3D materials or 1D grain boundaries in 2D materials, can be particularly damaging since they directly impede the transport of charge, spin or heat, and can introduce a metallic character into otherwise semiconducting systems. Unfortunately, a technique to rapidly and non-destructively image 1D defects in 2D materials is lacking. Scanning transmission electron microscopy (STEM)~\cite{hone2013,lou2013,lou2014,nasim2014}, Raman~\cite{cancado2011,mignuzzi2015}, photoluminescence~\cite{carozo2017,alencar2016,hone2013,lou2014} and nonlinear optical spectroscopies~\cite{Yin2014,karvonen2017}, are all extremely valuable, but current implementations suffer from low throughput and a destructive nature (STEM) or limitations in their unambiguous sensitivity at the nanoscale. Here we demonstrate that dark-field second harmonic generation (SHG) microscopy can rapidly, efficiently, and non-destructively probe grain boundaries and edges in monolayer dichalcogenides (i.e. MoSe$_2$, MoS$_2$ and WS$_2$). Dark-field SHG efficiently separates the spatial components of the emitted light and exploits interference effects from crystal domains of different orientations to localize grain boundaries and edges as very bright 1D patterns through a \v{C}erenkov-type SHG emission. The frequency dependence of this emission in MoSe$_2$ monolayers is explained in terms of plasmon-enhanced SHG related to the defect's metallic character. This new technique for nanometer-scale imaging of the grain structure, domain orientation and localized 1D plasmons in 2D different semiconductors, thus enables more rapid progress towards both applications and fundamental materials discoveries.
}}

Atomically thin 2D semiconductor transition metal dichalcogenides (TMDs) host a diversity of excitons with robust light-valley coupling and exciton-defect coupling that allows spectroscopic characterization~\cite{heinzreview2014, mak2016, lin2016defect}. Different techniques have been employed to characterize 1D defects in these 2D systems to evaluate sample crystalline quality for potential device applications~\cite{lin2016defect}. Transmission electron microscopy directly resolves atomistic details of defects~\cite{Zhou2013,Ossi2015,hone2013,lou2013,lou2014,nasim2014}, but requires intensive and disruptive sample preparation. Photoluminescence and Raman spectroscopy provide rapid and non-destructive probes of the electronic and vibrational properties of defective regions~\cite{hone2013,lou2014,carozo2017,karvonen2017}, which often manifest as red- or blue-shifted emission with enhanced or suppressed intensities when compared to the response from a pristine crystal, depending on multiple factors that affect local electronic properties such as material composition, doping level, defect passivation~\cite{Nan2014}, grain boundary (GB) geometry~\cite{hone2013} or edge terminations~\cite{Lin2018}. This complicates their use as means of reproducible defect characterization. Therefore, a more reliable and facile method to effectively image grain boundaries and edges \textit{independent of composition, doping, and defect reconstruction}, is needed. Recent reports show that nonlinear optical spectroscopy is highly sensitive for imaging 1D imperfections~\cite{Yin2014,karvonen2017}; nonetheless, their visualization exhibited a weak background contrast~\cite{Yin2014} or required the use of chemical solvents to enhance image contrast~\cite{karvonen2017}.

Dark-field (DF) microscopy boost image contrast by suppressing light scattered from the homogeneous regions. The technique does not depend on the detailed local atomic and electronic structure, i.e. sharp discontinuities always appear brighter. When comparing a bright-field (BF) linear optical image of a monolayer MoS$_2$ on quartz (Supplementary Figure~S1a), and a DF linear optical image (i.e. incident and collected light of the same frequency, Supplementary Figure~S1b), we observe in DF light scattered from the edges that are transparent to BF microscopy. However, atomically thin grain boundaries cannot be resolved by either imaging technique, because the in-plane linear dielectric response of MoS$_2$ monolayers is isotropic so radiation dipoles generated by two grains separated by a GB are always in phase, making the scattered light intensity near a GB indistinguishable from that in the pristine region, even in dark field imaging. (The bright dots in the DF image are due to contaminants from the transfer process, see Methods section). In stark contrast, second harmonic radiation dipoles (emitting frequency-doubled photons) in two grains separated by a mirror boundary are always out of phase, thus creating interference that can be detected as a localized feature. Here we develop a dark-field nonlinear characterization technique combining SHG microscopy and a spatial filter, to further enhance the second harmonic contrast of 1D imperfections in 2D systems, thus allowing detailed large-area spatial mapping of grain boundaries and edges regardless of their local atomic and electronic structures.


Monolayer TMDs grown by chemical vapour deposition (see Methods) were characterized using a picosecond laser system tunable from 1.30 to 1.65~eV. SHG microscopy was performed using a high numerical aperture objective in a back-reflected geometry (details in Methods). We employed a DF filter (patch stop) in front of the detector to collect only light with large wave-vectors (see Supplementary Note~S2). 
%
Figures~\ref{fig1}a,b show BF- and DF-SHG images of monolayer MoSe$_2$ for a pump beam of 1.38~eV (900~nm), where the uniform SHG intensities in interior of each grain confirms their single-layer nature~\cite{Yin2014,karvonen2017}. A dark GB in Fig.~\ref{fig1}a is ascribed to destructive interference of the SHG signals generated from neighboring grains with opposite orientations~\cite{Yin2014,karvonen2017}. The DF-SHG response (Fig.~\ref{fig1}b) of the central region is considerably suppressed as a result of the DF filter blocking SHG emission from the bulk, leaving GBs and edges now as bright emission features.

To estimate the contrast difference between BF- and DF-SHG, line-scans (dashed white line) are performed along the grain boundaries, edges and bulk regions (insets Figs.~\ref{fig1}a,b); normalized by the bulk signal. We define a 1D-defect contrast factor as $\text{C}= \frac{I_{\text{1D}}-I_{\text{bulk}}}{I_{\text{1D}}+I_{\text{bulk}}}$, where $I_{\text{1D}}$ and $I_{\text{bulk}}$ are the SHG responses of the GB/edge and the bulk; this contrast can be calculated for both dark ($\text{C}_{\text{DF}}$) and bright ($\text{C}_{\text{BF}}$) fields. The line profiles of Figs.~\ref{fig1}a,b yield a three-fold contrast enhancement from the grain boundary under dark-field conditions, ${\left| \frac{\text{C}_{\text{DF}}}{\text{C}_{\text{BF}}}\right|_{\text{GB}}}=3.0\pm0.3$, and a dramatic 20-fold enhancement from the edge, ${\left| \frac{\text{C}_{\text{DF}}}{\text{C}_{\text{BF}}}\right|_{\text{edge}}}=20\pm5$. To establish the potential of this technique, Figure~\ref{fig2} shows DF-SHG imaging for several additional monolayers of MoSe$_2$, MoS$_2$ and WS$_2$. All show strong contrast enhancements, with some variations arising from variations in sample size and optimization of the patch stop position. These results show the great potential of DF-SHG as a novel highly sensitive, non-invasive probe to effectively reveal structural discontinuities in 2D semiconductors without the use of chemical solvents, with fast image rates and high spatial resolutions compared to other optical spectroscopy techniques when imaging the same type of defects~\cite{hone2013,lou2014,Yin2014,carozo2017,karvonen2017}  (Supplementary Movies~S1 and~S2 show a fast and large area DF-SHG scan).

To explain the DF-SHG process near a MoSe$_2$ GB, we first determine its atomic structure using aberration-corrected high-resolution scanning transmission electron microscopy (AC-HRSTEM). Figure~\ref{fig1}c shows a low-magnification STEM image of monolayer MoSe$_2$. A high-angle annular dark field (HAADF) detector was used for STEM imaging and the edge of the MoSe$_2$ flake is highlighted by a yellow dotted line. Electron diffraction (Fig.~\ref{fig1}d) acquired from the area circled in green in Fig.~\ref{fig1}c shows a pair of first-order hexagonal diffraction spots. Placing an objective aperture at selected diffraction spots yields DF-TEM images (Supplementary Figure~S3) that are overlaid with the original STEM-HAADF image (Fig.~\ref{fig1}c) to reveal two mirror twin grains with opposite orientations, colored in red or blue in Fig.~\ref{fig1}d. Detailed atomic images of GBs include the 4$\mid$4 motifs aligned along the zigzag direction (Figure~\ref{fig1}e,f), and 4$\mid$8 reconstructions that introduce tilts and kinks (Figure~\ref{fig1}g and Supplementary Figure~S4)~\cite{Zhou2013,Ossi2015}.

With the atomic grain structure determined, we now discuss the mechanism behind the increased contrast of 1D defects in the DF-SHG images. Figures~\ref{fig3}a,b shows a second harmonic far-field spatial mode that results when the pump beam (at 950~nm) is focused on the center of the MoSe$_{2}$ monolayer (2D bulk) and GB, respectively. The far-field second harmonic pattern of the 2D bulk (green circle) preserves the circular symmetry of the Gaussian distribution of the incident light. In contrast, the second harmonic emission from the GB has a dark central stripe with additional arc features at large angles as shown in Fig.~\ref{fig3}b. Since mirror grains have opposite polarities ($x \rightarrow -x$) and switch the sign of the  second-order nonlinear susceptibility $\chi^\text{(2)}_{xxx}$, the out-of-phase second harmonic electric field radiated from mirror grains will interfere destructively along small angles $\theta$ (with respect to the normal of the sample plane, inset in Figure~\ref{fig3}), as can be described within a dipole model (see Supplementary Note~S4). 
This explains the dark stripe in Fig.~\ref{fig3}b~\cite{Yin2014}. On the other hand, larger $\theta$ can compensate for the phase offset from the two mirror grains, as constructive interference following $d \sin\theta = \lambda/2$, where $d$ is the distance between the two grains within a laser spot size (radius of the Gaussian profile), and $\lambda$ the SHG wavelength ($\lambda=475$~nm). Figure~\ref{fig3}c depicts this situation, where two anti-parallel domains give rise to a second harmonic emission that is directed at larger angles $\theta$, as shown by the calculated 3D surface radiation (Fig.~\ref{fig3}c). Therefore, the SHG response from GB regions is emitted at larger angles, yielding the arc features shown in Fig.~\ref{fig3}b and calculated in Fig.~\ref{fig3}c. The effective spot size corresponds to $d=400$~nm, as obtained from the SHG image analysis considering a Gaussian profile. We can thus estimate the direction of constructive interference to be at $\theta \approx (36 \pm 2)^\circ$, close to the 32$^\circ$ measured from the far-field pattern (Fig.~\ref{fig3}b and Supplementary Note~S5 for angle determination). 

A similar anisotropy in second harmonic radiation has been observed for inversion domain boundaries in bulk nonlinear crystals, where due to the large thickness of the crystals phase matching can also occur, leading to so-called \v{C}erenkov phase-matched second harmonic emission~\cite{tunyagi03,sheng10}. In fact, any sharp discontinuity in $\chi^{(2)}$ leads to this type of anisotropic second harmonic radiation. In the present case, the symmetry of the GB requires that $+\chi^{(2)} \rightarrow -\chi^{(2)}$ and the sample's edges require $\chi^{(2)} \rightarrow 0$ due to the termination of nonlinear material~\cite{sheng12,ren13,roppo13,zhao16}. In our DF-SHG experiment, the patch stop removes the SHG radiation emitted at small angles, thereby favoring emission from 1D GBs and edges in semiconducting TMDs (Figs.~\ref{fig1},~\ref{fig2}).

We next explore the GB DF-SHG signal for different fundamental photon energies. Figure~\ref{fig4}a shows the DF-SHG amplitude dependence on the photon energy for bulk (red squares) and GB (green spheres) from monolayer MoSe$_{2}$ from 1.32 to 1.65~eV (Supplementary Note~S6 shows similar data from an edge). The DF-SHG bulk emission shows an onset of optical transitions at $\sim$1.55~eV, which has been attributed to a resonance close to the $A$ exciton energy~\cite{kim2015}, where the energy of one of the two photons equals the optical gap (so-called $\omega$ resonance). For excitations between 1.32 and 1.51~eV, the weak resonance near 1.3~eV for the bulk emission can be associated to a two-photon resonance ($2\omega$ resonance), where twice the photon energy matches the C exciton energy or other low-energy states due to structural defects present within the bulk crystal lattice~\cite{louie2018}. By contrast, the DF-SHG GB emission shows a strong resonance at 1.32~eV compared to the bulk case, yielding an enhanced contrast between the two signals.

Constructive interference alone would yield a near-constant intensity ratio between the GB and the bulk emission as a function of frequency, which is not the case here. Thus, the strong frequency dependence shown in Fig.~\ref{fig4}a indicates the presence of additional SHG-modulating processes. Similar frequency-dependent BF-SHG intensities from 1D defects in MoS$_2$ have been reported \cite{Lin2018,Yin2014} where the measured edge SHG enhancement reaches 1.4$\times$ at half the optical gap of MoS$_2$. If a $\sim 1$ nm wide edge region (effective edge thickness taken in Ref.~\citenum{Yin2014}) within a 1~$\mu$m laser spot amplifies the SHG by 1.4 times, the amplification factor intrinsic to the edge would be about 400 (Supplementary Note~S7 gives more details).

The \textit{selectivity} of the SHG contrast enhancement to the spectral range below the MoSe$_2$ optical gap of 1.5~eV, and the order-of-magnitude amplification inferred above from previous work~\cite{Lin2018,Yin2014}, are characteristic of plasmon resonances: strongly localized electric fields confined within 1D nanowires can amplify the optical response of the host material~\cite{Goodfellow2014}. Momentum conservation in such plasmon-mediated SHG processes can be achieved by the annihilation of two counter-propagating plasmons in 2D~\cite{Chen1979,Simon1974} or coupled 1D+2D systems~\cite{Li2017}. Plasmons in MoSe$_2$ samples originate from metallic states localized at mirror grain boundaries (Supplementary Note~S7). To show that plasmon enhancement in SHG are limited to $<$1.5~eV, we calculate the energy range in which the plasmon localized on the GB carries a significant spectral weight, as can be characterized by the calculated mode-specific electron energy loss spectra (EELS). Although photons cannot provide the large wave-vectors needed to excite plasmons localized along a pristine GB and although the longitudinal dielectric response is calculated here (accessible to EELS but not to optical spectroscopy), the breaking of translational symmetry by kinks along the GB (Fig.~\ref{fig1}g) could supply the required momentum and allow photon-plasmon coupling, similar to that seen in bent nanowires~\cite{Sanders2006} and finite-length 1D metallic quantum wires~\cite{Nagao2010,Chung2010}. This assumption is consistent with kink patterns observed in STEM (Fig.~\ref{fig1}g and Supplementary Figure~S4). Thus, we only consider the calculated EELS to reveal available plasmonic states in an ideal 1D GB, and assume the presence of sufficient translational symmetry breaking to activate plasmon-photon coupling.
 
The calculated mode-specific EELS, $-\Im[\epsilon_n(q,\omega)]$, is shown in Figure~\ref{fig4}b, where we select the mode index $n$ for the plasmon localized on the 4$\mid$4 GB (Supplementary Note~S7). As the momentum transfer increases from 0 to 0.6 $\pi/a$, where $a$ is the MoSe$_2$ lattice constant 3.29~\AA, the spectral weight of the plasmon localized on the 4$\mid$4 GB decreases towards zero as the plasmon peak energy approaches 1.5~eV. This plasmon dispersion is shown in Figure~\ref{fig4}c (solid curve), where the circle sizes are proportional to plasmon spectral weights. The dispersion for intra-band transition resonances, $\Im(\epsilon_n(\omega))$, are plotted in a dashed line. 
Since each plasmon peak is determined by
$\Re(\epsilon_n(\omega)) \equiv \epsilon_r =0$, the decrease in the EELS peak spectral weight, $-\Im(\epsilon{_n}(\omega))={\left. {\frac{\epsilon{_i}}{\epsilon{_r^2}+\epsilon{_i^2}}} \right|_{({\text{EELS peak}})}} = \frac{1}{\epsilon{_i}} $, means that the imaginary part of the dielectric function is increasing due to the increasing contributions from inter-band transitions as the excitation energy approaches the band gap, i.e. plasmons become damped by inter-band transitions approaching the 1.5~eV gap~\cite{Andersen2014}. The calculated existence of a 1D plasmon mode and its complete damping beyond 1.5~eV are consistent with the measured SHG amplification only below 1.5~eV, as shown in Fig.~\ref{fig4}a.


In conclusion, dark-field SHG microscopy provides rapid, high-contrast, non-destructive mapping of 1D defects in atomically thin 2D TMDs due to an interference-enhanced directional confinement of the nonlinear response across the 1D defect in a manner analogous to the \v{C}erenkov phase-matched second harmonic emission. The observed frequency-dependent DF-SHG enhancement near the grain boundary suggests a prominent plasmon-modulated nonlinear optical response at 1D defects in atomically thin TMDs. Our results provide a direct pathway to electronically map the SHG response of 1D defects acting as nanoantennae within a 2D layered TMD, potentially applicable to other classes of 2D systems.

\section*{Methods}


Methods, including statements of data availability and any associated accession codes and references, are available at Supplementary Information.

\paragraph*{Data availability.} All relevant data are available from the authors upon request.

\section*{Acknowledgments} 
B.R.C., A.M.dP., M.A.P., and L.M.M. acknowledge the financial support from the Brazilian agencies CNPq, CAPES, FAPEMIG and Brazilian Institute of Science and Technology of Nanocarbon (INCT-Nanocarbono) and Molecular Medicine (INCT-Medicina Molecular). 
Y.W. and V.H.C acknowledge support from the National Science Foundation 2-Dimensional Crystal Consortium - Materials Innovation Platform (2DCC-MIP) under DMR-1539916.
K.F., P.M.A. and M.T. acknowledge The Air Force Office of Scientific Research (AFOSR) grant 17RT0244 for financial support. T.Z. and M. T. also thank the financial support from the National Science Foundation (2DARE-EFRI-1433311). S.K. acknowledges financial support from NSF ECCS 1351424.
%
\section*{Author contributions}
B.R.C. and L.M.M. conceived the idea and designed the experiments. B.R.C. and L.M.M. performed the nonlinear optical experiments. 
A.M.dP. contributed to the nonlinear experiment in its early phase.
B.R.C. and M.A.P. performed the Raman experiments. B.R.C. analyzed and interpreted the nonlinear and Raman experimental data. Y.W. and V.H.C. performed the theoretical work.
K.F., T.Z., E.K. and M.T. performed the scanning transmission electron measurements and synthesized the MoS$_2$ and WS$_2$ monolayers.
I.B., S.K. and P.M.A. synthesized the MoSe$_2$ flakes.
B.R.C., Y.W. and L.M.M. wrote the manuscript. All the authors discussed the results and commented on the manuscript.
\paragraph*{Competing interests}
The authors declare no competing financial interests.

\section*{Additional information}
\paragraph*{Supplementary Information} accompanies this paper at 
\paragraph*{Correspondence and requests for materials} should be addressed to B.R.C. or L.M.M.


\bibliographystyle{ieeetr}
\bibliography{references}
%


\begin{figure}[!htbp]
    \centering
    \includegraphics[width=0.7\textwidth]{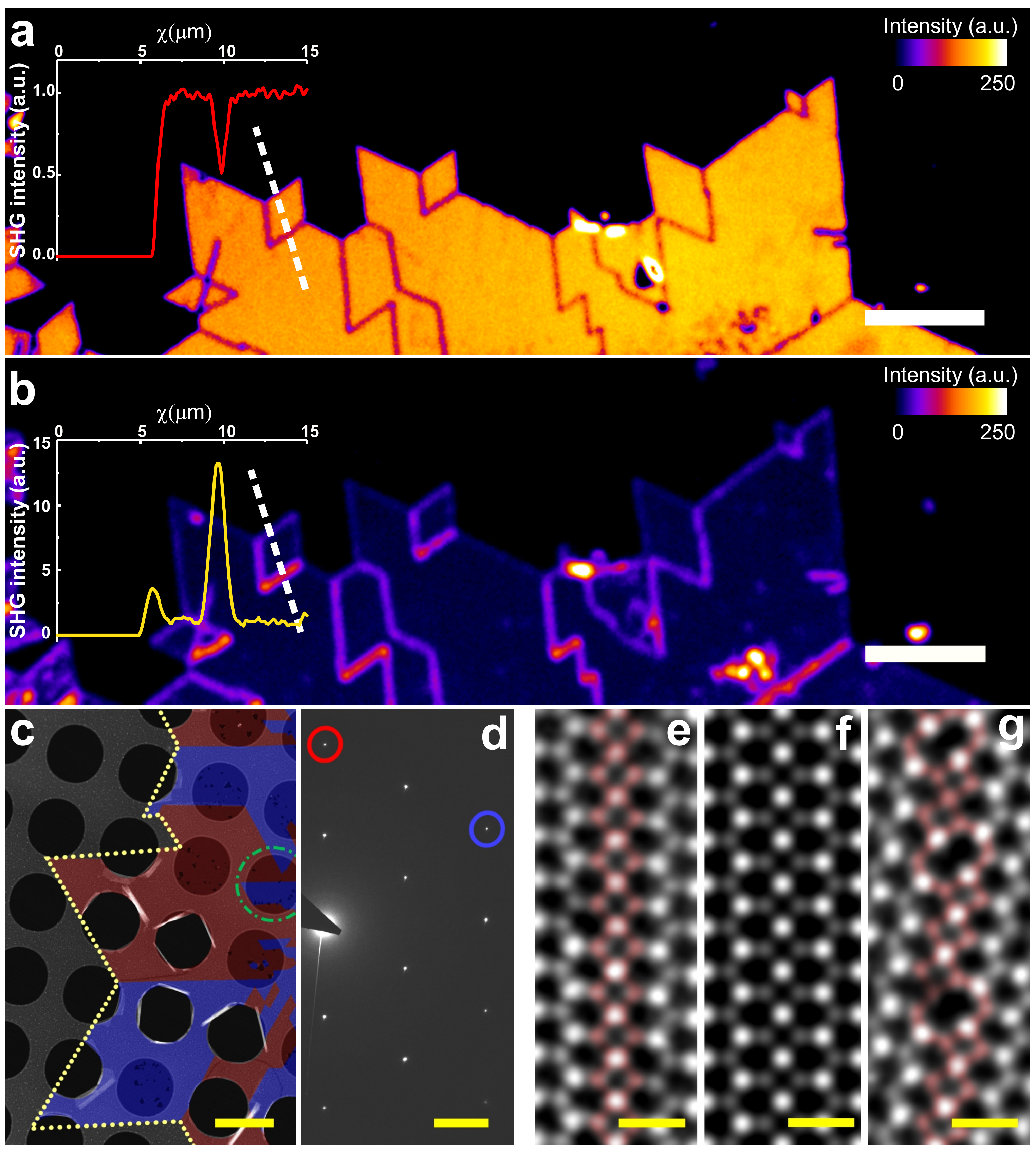} 
    \caption{{\bf Bright grain boundaries emission obtained by dark-field second harmonic generation (DF-SHG).}
    {\bf a.} Bright- and {\bf b.} dark-field SHG imaging of a CVD-grown monolayer MoSe$_2$ pumped at 1.38~eV, scale bars are 10~$\mu$m. In ({\bf b}) a bright grain boundary and a slight edge enhancement are clearly revealed by the DF-SHG intensity. The inset in ({\bf a}) and ({\bf b}) depict the cross-section (dashed white line), used to compare the contrast of the BF- and DF-SHG response on the same crystal.
    {\bf c.} False-colored STEM-HAADF image of monolayer MoSe$_2$ and 
    {\bf d.} electron diffraction pattern acquired from green-circled area of {\bf c}. Mirror-twin-domains are identified by dark-field (DF)-TEM technique using red- and blue-circled diffraction spots in {\bf c}, and the HAADF image was colored in red and blue, corresponding to different orientations of mirror-twins. The red (blue)-colored area show higher DF-TEM signal when the red (blue)-circled diffraction spot was selected with an objective aperture. Scale bars in ({\bf c}, {\bf d}) correspond to 2~$\mu$m and 2 nm$^{-1}$, respectively. 
    {\bf e}--{\bf g.} Atomic-resolution STEM-HAADF images show 4$\mid$4 and 4$\mid$8 configurations of mirror twin grain boundary. Experimentally acquired image ({\bf e}) and simulated STEM image ({\bf f}) of 4$\mid$4 GB, and experimentally acquired image ({\bf g}) of 4$\mid$8 GB. Scale bars in ({\bf e}--{\bf g}) correspond to 500~pm.
}
    \label{fig1}
\end{figure}

\begin{figure}[!htbp]
    \centering
    \includegraphics[width=1.0\textwidth]{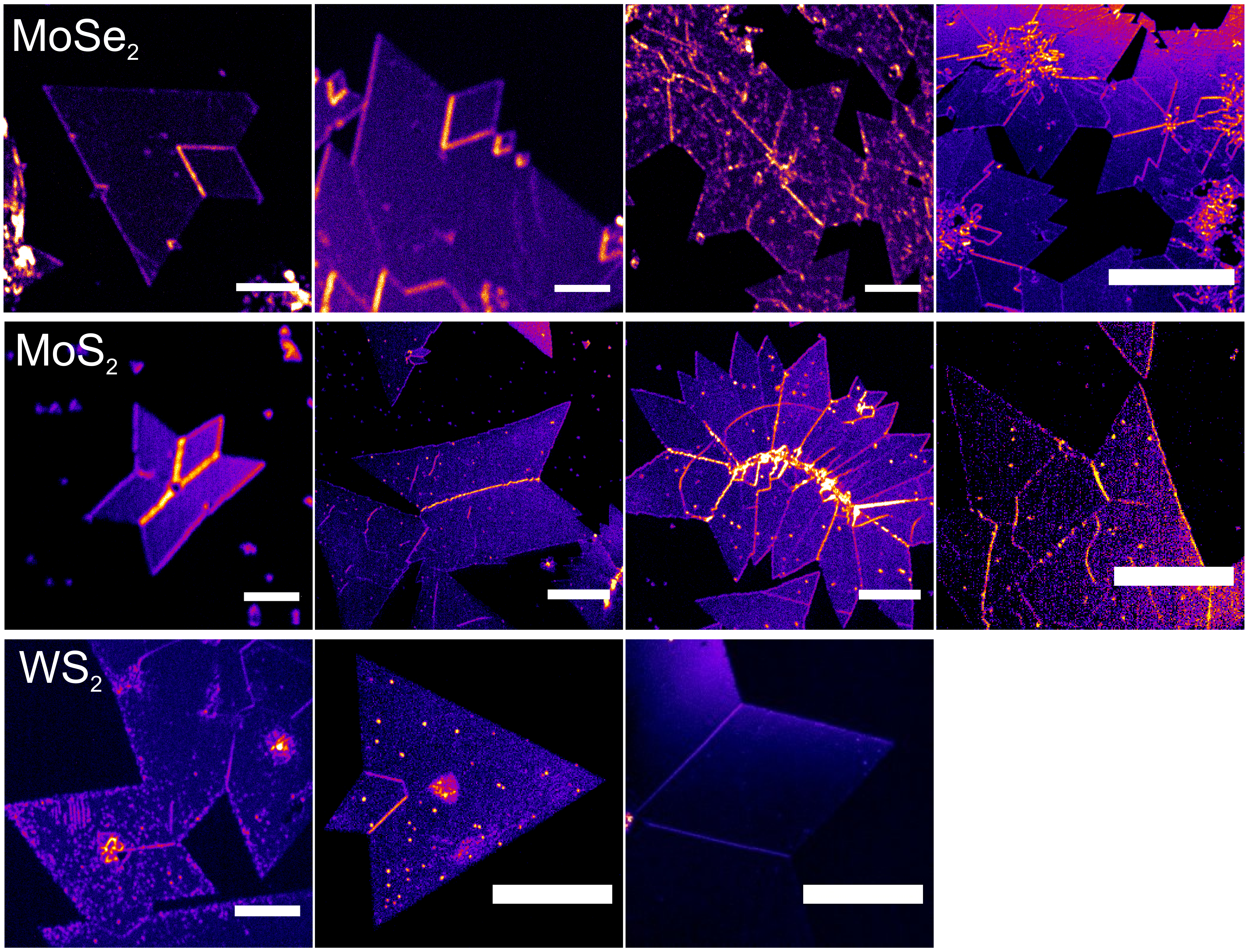} 
    \caption{{\bf Characterization of 1D objects in semiconducting TMDs.} 
Dark-field second harmonic generation imaging on different MoSe$_2$ (top row), MoS$_2$ (middle row) and WS$_2$ (bottom row) monolayers on quartz substrates collected by a microscope objective of 40$\times$ (N.A.=0.95). The samples were excited with an energy of 1.38~eV for MoSe$_2$ and 1.42~eV for MoS$_2$ and WS$_2$, respectively. Scale bars correspond to 20~$\mu$m for MoSe$_2$ and 10~$\mu$m for MoS$_2$ and WS$_2$.
}
    \label{fig2}
\end{figure}

\begin{figure}[!htbp]
    \centering
    \includegraphics[width=1.0\textwidth]{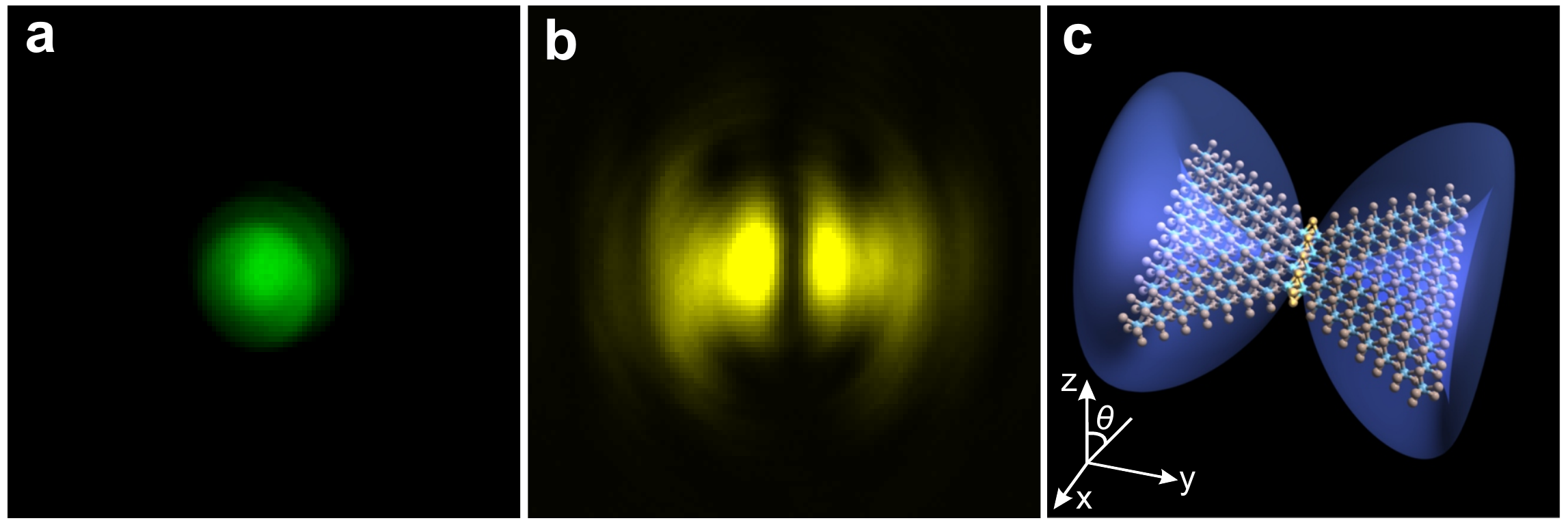} 
    \caption{{\bf Far-field spatial mode pattern.} 
    Far-field images of a CVD as-grown monolayer MoSe$_2$ produced when the pumped beam is focused on {\bf a.} bulk monolayer and {\bf b.} grain boundary.
    {\bf c.} Sketch of the MoSe$_{2}$'s two domains with opposite second harmonic phase separated by the grain boundary (middle) giving rise to a second harmonic emission pattern with large angles (blue conical shape).
}
    \label{fig3}
\end{figure}

\begin{figure}[!htbp]
    \centering
    \includegraphics[width=0.9\textwidth]{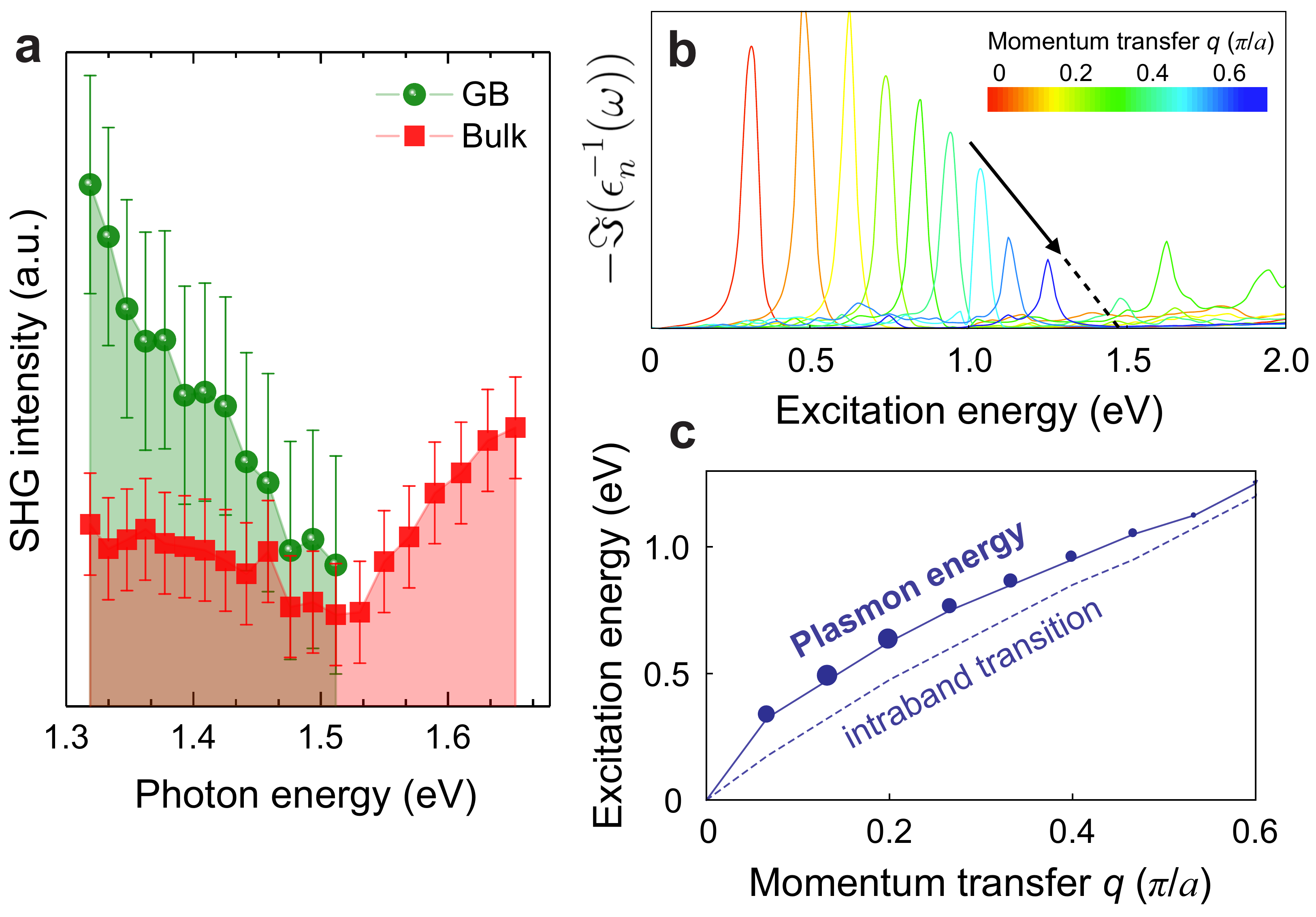} 
    \caption{{\bf Plasmon resonance on a 1D defect in MoSe$_2$ monolayer.} 
    {\bf a.} DF-SHG excitation profiles of GB and bulk for monolayer MoSe$_2$ on quartz substrate. The data were normalized by the quartz intensity of each excitation energy.
    {\bf b.} Calculated electron energy loss spectra for a mirror 4$\mid$4 grain boundary in MoSe$_2$, where the plasmon momentum varies from 0 to 0.6 $\pi/a$.
    The arrow in ({\bf b}) highlights the 1.5 eV cutoff energy.
    {\bf c.} Calculated dispersion for a 1D plasmon along a mirror grain boundary (solid curve) and dispersion for intraband transition resonances (dashed line).
}
    \label{fig4}
\end{figure}

\end{document}